\documentclass[11pt]{article}
\usepackage{amsmath}
\usepackage{amsfonts}
\usepackage{amssymb}
\usepackage{graphicx}
\usepackage{bm}
\usepackage{color}
\usepackage{amscd}

\def\bea{\begin{eqnarray}}
\def\eea{\end{eqnarray}}

\begin{document}
\begin{center}
\LARGE {\bf  Quasi-local conserved charges in the Einstein-Maxwell theory }
\end{center}

\begin{center}
{M. R. Setare \footnote{E-mail: rezakord@ipm.ir}\hspace{1mm} ,
H. Adami \footnote{E-mail: hamed.adami@yahoo.com}\hspace{1.5mm} \\
{\small {\em  Department of Science, University of Kurdistan, Sanandaj, Iran.}}}\\

\end{center}

\begin{center}
{\bf{Abstract}}\\
 In this paper we consider the Einstein-Maxwell theory and define a combined transformation composed of diffeomorphism and $U(1)$ gauge transformation. For generality, we assume that the generator $\chi$ of such transformation is field-dependent. We define the extended off-shell ADT current and then off-shell ADT charge such that they are conserved off-shell for the asymptotically field-dependent symmetry generator $\chi$. Then, we define the conserved charge corresponding to the asymptotically field-dependent symmetry generator $\chi$. We apply the presented method to find the conserved charges of the asymptotically AdS$_{3}$ spacetimes in the context of the Einstein-Maxwell theory in three dimensions.  Although the usual  proposal for the quasi local charges provides divergent global charges for the Einstein-Maxwell theory with negative cosmological constant in three dimensions, here we avoid this problem by introducing proposed combined transformation $\chi$.
\end{center}

\section{Introduction}
The concept of conserved charges is a very important matter
in gravity theories as well as in other physical theories.
As is well known, the concept of conserved charges of gravity
theories is related to the concept of the Noether charges
corresponding to the Killing vectors which are admitted by
solutions of a theory. A method to calculate the energy of the asymptotically AdS solution was given by Abbott and Deser \cite{9'}. Deser and Tekin have extended this approach to the calculation of the energy of the asymptotically dS or AdS solutions in higher curvature gravity models \cite{10'}.  The authors of \cite{1} have obtained the quasi-local conserved charges for  black holes in any diffeomorphically invariant theory of gravity. By considering an appropriate variation of the metric, they have established a one-to-one correspondence between the Abbott, Deser and Tekin (ADT) approach and the linear Noether expressions. They have extended this work to a theory of gravity containing a gravitational Chern-Simons term in \cite{17}. In this paper we are going to obtain the quasi-local conserved charges of the Einstein-Maxwell theory. Recently the authors of \cite{10} have studied the asymptotic structures of AdS spacetimes of the Einstein-Maxwell theory in 3 dimensions.
 Asymptotic symmetry was  applied with
success  some time ago  to  asymptotically 3D anti-de Sitter
(AdS$_{3}$) spacetimes, to show that the
asymptotic symmetry  group (ASG) of AdS$_{3}$ is the
conformal group in two dimensions \cite{1'}. This fact
represents the first evidence of the existence of an anti-de
Sitter/conformal field theory (AdS/CFT)
correspondence and was later used by Strominger to explain the
Bekenstein-Hawking entropy of the Ba$\tilde{\text{n}}$ados, Teitelboim and Zanelli (BTZ) black hole in terms of the
degeneracy of states of the boundary CFT generated by the asymptotic
metric deformations \cite{2'}. In order to determine the ASG, one has first to fix boundary
conditions for the fields at $r=\infty$ then to find the Killing
vectors leaving these boundary conditions invariant.
The boundary conditions  must be relaxed enough to allow for the
action of the conformal group and for the right boundary
deformations, but tight enough to keep finite the charges associated
with the ASG generators, which are given by boundary terms of the
action. The authors of \cite{10} have shown that, for a generic choice of boundary conditions, the asymptotic symmetries of the Einstein-Maxwell theory in 3 dimension are broken down to $R\otimes U(1)\otimes U(1)$. Here we define a combined transformation composed of diffeomorphism and $U(1)$ gauge transformation and assume that the generator $\chi$ of such transformation is field-dependent. Then we define the extended off-shell ADT current which is conserved off-shell for the asymptotically field dependent symmetry generator $\chi$. Using this definition we obtain the extended off-shell ADT charge. By integrating the extended off-shell ADT charge over a spacelike codimension-2 surface, we obtain the conserved charge perturbation corresponding to the asymptotically field dependent symmetry generator $\chi$. As an example, we apply our method to find conserved charges of the asymptotically AdS$_{3}$ spacetimes in the context of the Einstein-Maxwell theory in 3 dimensions. Our results for the conserved charge corresponding to the pure $U(1)$ gauge symmetry and for the conserved charge corresponding to the asymptotically Killing vector $\xi$, are consistent with those of \cite{10}, where the authors used the Hamiltonian formalism to find the corresponding results.
\section{Quasi-local conserved charges in the Einstein-Maxwell theory}
The quasi-local method for finding conserved charges in the covariant theory of gravity have presented in the paper \cite{1}, where the conserved charges corresponding to the Killing vectors admitted by spacetime everywhere. The authors of \cite{2} have generalized this approach such that contains the asymptotically Killing vectors as well as the Killing vectors admitted by spacetime everywhere. The method presented in \cite{1}, which is valid for covariant theory of gravity, have been extended to the covariant theory of gravity coupled to matter fields in \cite{3}. Here we will go to generalize the approach presented in \cite{3} such that it becomes suitable to calculate the conserved charges in the context of the Einstein-Maxwell theory, correspond to field-dependent (asymptotically) Killing vector fields as well as field-independent one.\\
The Lagrangian density of the Einstein-Maxwell theory is a functional of metric $g_{\mu \nu}$ and the gauge field $A_{\mu}$,
\begin{equation}\label{1}
  L= \sqrt{-g} \mathcal{L}(g_{\mu \nu}, A_{\mu}),
\end{equation}
where
\begin{equation}\label{2}
  \mathcal{L}= R-2 \Lambda - \frac{\kappa}{2} F_{\mu \nu} F^{\mu \nu}
\end{equation}
 here $R$, $F_{\mu \nu} = \partial _{\mu} A_{\nu} - \partial _{\nu} A_{\mu} $, $\Lambda$ are respectively the Ricci scalar, field strength, the cosmological constant and $\kappa = 8 \pi G$. In this theory a $U(1)$ gauge field $A_{\mu}$ is minimally coupled to gravity and we can write the Lagrangian Eq.\eqref{1} as $L=L_{(g)}+ L_{(A)}$. The variation of the Lagrangian Eq.\eqref{1} with respect to $g_{\mu \nu}$ and $A_{\mu}$ is
\begin{equation}\label{3}
  \delta L = \sqrt{-g} \left( \mathcal{E}_{(g)}^{\mu \nu} \delta g_{\mu \nu} + \mathcal{E}_{(A)}^{\mu} \delta A_{\mu} \right)+ \partial_{\mu} \Theta^{\mu}(\Phi, \delta \Phi),
\end{equation}
where $\Phi = \{g_{\mu \nu},A_{\mu}\}$. In the equation \eqref{3}, $\mathcal{E}_{(g)}^{\mu \nu} = \mathcal{E}_{(A)}^{\mu} =0$ are the equations of motion and $\Theta^{\mu}(\Phi, \delta \Phi)$ is just surface term, which are given as
\begin{equation}\label{4}
  \mathcal{E}_{(g)}^{\mu \nu}= - \left( G^{\mu \nu} + \Lambda g^{\mu \nu} \right)+ \kappa T^{\mu \nu},
\end{equation}
\begin{equation}\label{5}
  \mathcal{E}_{(A)}^{\mu}=2 \kappa \nabla _{\nu} F^{\nu \mu},
\end{equation}
\begin{equation}\label{6}
  \Theta^{\mu}(\Phi, \delta \Phi)= \Theta _{(g)}^{\mu}(\Phi, \delta \Phi) + \Theta _{(A)}^{\mu}(\Phi, \delta \Phi),
\end{equation}
with
\begin{equation}\label{7}
\begin{split}
    \Theta _{(g)}^{\mu}(\Phi, \delta \Phi)= & 2 \sqrt{-g} \nabla ^{[\alpha} \left( g^{\mu] \beta} \delta g_{\alpha \beta} \right), \\
     \Theta _{(A)}^{\mu}(\Phi, \delta \Phi) = & -2 \kappa \sqrt{-g} F^{\mu \nu} \delta A_{\nu}.
\end{split}
\end{equation}
Equation \eqref{4} is well-known as the Einstein field equation and in this equation, $G^{\mu \nu}$ is the Einstein tensor and $T^{\mu \nu }$ is the electromagnetic energy-momentum tensor
\begin{equation}\label{8}
  T^{\mu \nu} = F^{\mu \alpha} F^{\nu}_{\hspace{1.7 mm} \alpha} - \frac{1}{4} g^{\mu \nu} F^{\alpha \beta} F_{\alpha \beta},
\end{equation}
also, equations \eqref{5} and
\begin{equation}\label{9}
  \nabla_{[\lambda}F_{\mu \nu]}=0
\end{equation}
are Maxwell field equations in the curved spacetime. By using the Bianchi identity, $\nabla _{\mu} G^{\mu \nu} =0$ and Eq.\eqref{9}, one can easily find that
\begin{equation}\label{10}
  \nabla _{\mu} \mathcal{E}_{(g)}^{\mu \nu}= \kappa F^{\nu \alpha} \nabla ^{\beta} F_{\beta \alpha} .
\end{equation}
It is clear that the right hand side of the equation \eqref{10} vanishes on-shell, but here we are interested to work off-shell.\\
We consider a combined transformation of the diffeomorphism and the $U(1)$ gauge transformation and we assume that $\chi= (\xi, \lambda)$ is the generator of such transformations, where $\xi=\xi^{\mu}(x) \partial_{\mu}$ is a vector field and $\lambda= \lambda (x)$ is a scalar field. The metric and the $U(1)$ gauge field under transformation generated by $\chi$ transform as
\begin{equation}\label{11}
  \delta _{\chi} g_{\mu \nu} = \pounds_{\xi} g_{\mu \nu},
\end{equation}
\begin{equation}\label{12}
  \delta _{\chi} A_{\mu} = \pounds _{\xi} A_{\mu} + \partial _{\mu} \lambda,
\end{equation}
where $\pounds_{\xi}$ denotes the Lie derivative along the vector field $\xi$.
It is clear that under transformation generated by $\chi$ the Lagrangian Eq.\eqref{1} transforms as
\begin{equation}\label{14}
  \delta _{\chi} L = \pounds _{\xi} L = \partial _{\mu} \left( \xi ^{\mu} \sqrt{-g} \mathcal{L} \right).
\end{equation}
Now, we suppose that the variation in Eq.\eqref{3} is generated by $\chi$
\begin{equation}\label{15}
  \delta _{\chi} L = \sqrt{-g} \left( \mathcal{E}_{(g)}^{\mu \nu} \delta_{\chi} g_{\mu \nu} +
 \mathcal{E}_{(A)}^{\mu} \delta_{\chi} A_{\mu} \right)+ \partial_{\mu} \Theta^{\mu}(\Phi, \delta_{\chi} \Phi).
\end{equation}
By substituting equations \eqref{11}, \eqref{12} and \eqref{14} into Eq.\eqref{15}, we have
\begin{equation}\label{16}
\begin{split}
   \partial _{\mu} \left( \xi ^{\mu} \sqrt{-g} \mathcal{L} \right) = & \sqrt{-g} \left( \mathcal{E}_{(g)}^{\mu \nu} \pounds _{\xi} g_{\mu \nu} + \mathcal{E}_{(A)}^{\mu} \pounds _{\xi} A_{\mu} \right) \\
     & + \sqrt{-g} \mathcal{E}_{(A)}^{\mu} \partial _{\mu} \lambda+ \partial_{\mu} \Theta^{\mu}(\Phi, \delta_{\chi} \Phi).
\end{split}
\end{equation}
On the one hand, by using equations \eqref{10}, one can easily find that
\begin{equation}\label{17}
  \mathcal{E}_{(g)}^{\mu \nu} \pounds _{\xi} g_{\mu \nu} + \mathcal{E}_{(A)}^{\mu} \pounds _{\xi} A_{\mu} =
2 \nabla _{\mu} \left( \mathcal{E}_{(g)}^{\mu \nu} \xi_{\nu} \right) + \nabla_{\mu} \left( \mathcal{E}_{(A)}^{\mu}
 \xi^{\alpha} A_{\alpha}\right),
\end{equation}
and on the other hand, because $\nabla _{\mu } \nabla_{\nu} F^{\mu \nu} = 0$, we have
\begin{equation}\label{18}
  \mathcal{E}_{(A)}^{\mu} \partial _{\mu} \lambda = \nabla_{\mu} \left( \lambda \mathcal{E}_{(A)}^{\mu} \right),
\end{equation}
therefore, the equation \eqref{16} can be written as
\begin{equation}\label{19}
  \partial _{\mu} J^{\mu}=0 ,
\end{equation}
where
\begin{equation}\label{20}
\begin{split}
   J^{\mu} (\Phi,\chi) = & \Theta^{\mu}(\Phi, \delta_{\chi} \Phi) - \xi ^{\mu} \sqrt{-g} \mathcal{L} + 2 \sqrt{-g} \mathcal{E}_{(g)}^{\mu \nu} \xi_{\nu} \\
     & + \lambda \sqrt{-g} \mathcal{E}_{(A)}^{\mu} + \sqrt{-g} \mathcal{E}_{(A)}^{\mu} \left(\xi^{\alpha} A_{\alpha}
 \right)
\end{split}
\end{equation}
It is clear that $ J^{\mu} (\Phi,\chi) $ is an off-shell current density for any vector field $\xi$ and for any $U(1)$ symmetry generator $\lambda$. By virtue of the Poincare lemma, one can write $J^{\mu} = \partial_{\nu} K^{\nu \mu}$. By substituting Eq.\eqref{2}, Eq.\eqref{4}, Eq.\eqref{5} and Eq.\eqref{6} into the Eq.\eqref{20} one can find the following expression for $K^{\mu \nu}$,
\begin{equation}\label{21}
  K^{\mu \nu} (\Phi;\chi) = K_{(g)}^{\mu \nu}(\xi) + K_{(A)}^{\mu \nu}(\lambda),
\end{equation}
where
\begin{equation}\label{22}
  K_{(g)}^{\mu \nu}(\xi) = 2 \sqrt{-g} \nabla ^{[\mu} \xi^{\nu]}, \hspace{1 cm} K_{(A)}^{\mu \nu}(\lambda) = 2 \kappa \sqrt{-g} F^{\mu \nu} \lambda .
\end{equation}
To keep the generality of discussion, we assume that $\xi$ and $\lambda$ are functions of the dynamical fields and $\hat{\delta}$ denotes variation with respect to the dynamical fields. The variation of the surface term Eq.\eqref{6} due to $\lambda$ is
\begin{equation}\label{23}
\begin{split}
    \delta _{\lambda} \Theta^{\mu}(\Phi, \hat{\delta} \Phi) & = \delta _{\lambda} \Theta _{(A)}^{\mu}(\Phi, \hat{\delta} \Phi) \\
     & = \partial _{\nu} K_{(A)}^{\nu \mu}( \hat{\delta} \lambda) - \hat{\delta} \lambda \sqrt{-g} \mathcal{E}_{(A)}^{\mu} - \Theta _{(A)}^{\mu}(\Phi, \delta_{\hat{\delta} \lambda} \Phi)
\end{split}
\end{equation}
Since the variation of the surface term Eq.\eqref{6} due to $\chi$ is $\delta _{\chi} \Theta^{\mu} = \pounds_{\xi} \Theta^{\mu} + \delta _{\lambda} \Theta^{\mu}$, so $\delta _{\chi} \Theta^{\mu}$ could be simplified as
\begin{equation}\label{24}
  \begin{split}
     \delta _{\chi} \Theta^{\mu} (\Phi, \hat{\delta} \Phi) = &  \xi ^{\mu} \partial_{\nu} \Theta^{\nu} (\Phi, \hat{\delta} \Phi) - \hat{\delta} \lambda \sqrt{-g} \mathcal{E}_{(A)}^{\mu} - \Theta ^{\mu} (\Phi, \delta_{(0,\hat{\delta} \lambda)} \Phi)   \\
       & + \partial _{\nu} \left( K_{(A)}^{\nu \mu}( \hat{\delta} \lambda) + 2 \xi^{[\nu} \Theta^{\mu]} (\Phi, \hat{\delta} \Phi) \right).
  \end{split}
\end{equation}
By varying Eq.\eqref{20} with respect to the dynamical fields and using Eq.\eqref{24}, we have
$$ \partial _{\nu} \left( \hat{\delta} K^{\nu \mu}(\Phi ; \chi) - K^{\nu \mu}(\Phi ; \hat{\delta} \chi)- 2 \xi ^{[\nu} \Theta ^{\mu]}(\Phi ; \hat{\delta} \Phi ) \right)$$
$$ = \hat{\delta} \Theta ^{\mu}(\Phi ; \delta _{\chi} \Phi ) - \delta _{\chi} \Theta ^{\mu}(\Phi ; \hat{\delta} \Phi ) - \Theta ^{\mu}(\Phi ; \delta _{\hat{\delta} \chi} \Phi ) $$
\begin{equation}\label{25}
  + 2 \sqrt{-g} \left( J_{ADT(g)}^{\mu}(\Phi,\hat{\delta} \Phi ;\xi) + J_{ADT(A)}^{\mu}(\Phi,\hat{\delta} \Phi;\xi)+ \frac{1}{2} \hat{\delta} \mathcal{E}_{(A)} ^{\mu} \lambda + \frac{1}{2} g ^{\alpha \beta} \hat{\delta} g_{\alpha \beta} \mathcal{E}_{(A)} ^{\mu} \lambda \right)
\end{equation}
where $J_{ADT(g)}^{\mu}(\Phi,\hat{\delta} \Phi ;\xi)$ and $J_{ADT(A)}^{\mu}(\Phi,\hat{\delta} \Phi;\xi)$ are the contributions from the metric and the gauge field in the off-shell ADT current which correspond to the diffeomorphism part \cite{1,3}, and they are given by
\begin{equation}\label{26}
   J_{ADT(g)}^{\mu}(\Phi,\hat{\delta} \Phi ;\xi)=\hat{\delta} \mathcal{E}_{(g)} ^{\mu \nu} \xi_{\nu} + \mathcal{E}_{(g)} ^{\mu \nu} \hat{\delta} g_{\nu \lambda} \xi ^{\lambda} - \frac{1}{2} \xi ^{\mu} \mathcal{E}_{(g)} ^{\alpha \beta} \hat{\delta} g_{\alpha \beta} + \frac{1}{2} g ^{\alpha \beta} \hat{\delta} g_{\alpha \beta} \mathcal{E}_{(g)}^{\mu \nu} \xi _{\nu},
\end{equation}
and
\begin{equation}\label{27}
  J_{ADT(A)}^{\mu}(\Phi,\hat{\delta} \Phi;\xi)=-\frac{1}{2} \xi ^{\mu} \mathcal{E}_{(A)}^{\nu}
 \hat{\delta} A_{\nu} + \left( \frac{1}{2} \hat{\delta} \mathcal{E}_{(A)} ^{\mu} + \frac{1}{2}
 g ^{\alpha \beta} \hat{\delta} g_{\alpha \beta} \mathcal{E}_{(A)} ^{\mu} \right)\xi^{\sigma}A_{\sigma},
\end{equation}
respectively. It is sensible definition of $J_{ADTU}^{\mu}(A;\lambda)$ as the contribution from the gauge field in the off-shell ADT current corresponds to the $U(1)$ transformation part
\begin{equation}\label{28}
  J_{ADTU(A)}^{\mu}(\Phi,\hat{\delta} \Phi;\lambda)= \frac{1}{2} \hat{\delta} \mathcal{E}_{(A)} ^{\mu} \lambda + \frac{1}{2} g ^{\alpha \beta} \hat{\delta} g_{\alpha \beta} \mathcal{E}_{(A)} ^{\mu} \lambda.
\end{equation}
Therefore, the off-shell ADT current which corresponds to $\chi$ can be defined as
\begin{equation}\label{29}
  \mathcal{J}_{ADT} ^{\mu}(\Phi,\hat{\delta} \Phi;\chi)=J_{ADT(g)}^{\mu}(\Phi,\hat{\delta} \Phi ;\xi) + J_{ADT(A)}^{\mu}(\Phi,\hat{\delta} \Phi;\xi)+J_{ADTU(A)}^{\mu}(\Phi,\hat{\delta} \Phi;\lambda).
\end{equation}
The off-shell ADT current $\mathcal{J}_{ADT} ^{\mu}(\Phi,\hat{\delta} \Phi;\chi)$ is conserved off-shell for arbitrary field-dependent Killing vector field which is admitted by the spacetime everywhere and for field-dependent $U(1)$ symmetry generator. Also, the symplectic current define as an antisymmetric bilinear map on perturbations \cite{8}
\begin{equation}\label{30}
  \omega ^{\mu} (\Phi; \delta _{1} \Phi , \delta _{2} \Phi) = \delta _{1} \Theta ^{\mu}(\Phi ; \delta _{2} \Phi ) - \delta _{2} \Theta ^{\mu}(\Phi ; \delta _{1} \Phi ) - \Theta ^{\mu}(\Phi ; [\delta _{1},\delta _{2}] \Phi ).
\end{equation}
The above expression for the symplectic current reduces to the Lee-Wald one \cite{4,5,6,7}, namely $\omega ^{\mu}_{\text{LW}} = \delta _{1} \Theta ^{\mu}(\Phi ; \delta _{2} \Phi ) - \delta _{2} \Theta ^{\mu}(\Phi ; \delta _{1} \Phi ) $ when two variations $\delta_{1}$ and $\delta_{2}$ are commute, i.e. $[\delta _{1},\delta _{2}] \Phi=0$. The symplectic current \eqref{30} is conserved on-shell and it gives us the conserved charges corresponding to the asymptotically field-dependent Killing vectors and for asymptotically field-dependent $U(1)$ symmetry generator. It should be noted that for the case in which $\chi$ is field-dependent we have $ [\hat{\delta},\delta _{\chi}] = \delta _{\hat{\delta} \chi} $, then Eq.\eqref{30} becomes
\begin{equation}\label{31}
  \omega ^{\mu} (\Phi; \hat{\delta} \Phi , \delta _{\chi} \Phi) = \hat{\delta} \Theta ^{\mu}(\Phi ; \delta _{\chi} \Phi ) - \delta _{\chi} \Theta ^{\mu}(\Phi ; \hat{\delta} \Phi ) - \Theta ^{\mu}(\Phi ; \delta _{\hat{\delta} \chi} \Phi ).
\end{equation}
It is easy to see that Eq.\eqref{31} reduces to the Lee-Wald symplectic current when $\chi$ is field-independent, i.e. $ \hat{\delta} \chi =0$. In the paper \cite{2}, the authors have generalized the off-shell ADT current in a generally covariant theory of gravity such that the generalized ADT current,
\begin{equation}\label{32}
   \mathcal{J}^{\mu} _{\text{GADT}} (g , \delta g ; \xi) =  \mathcal{J}^{\mu} _{\text{ADT}} (g , \delta g ; \xi) + \frac{1}{2 \sqrt{-g}} \omega ^{\mu} _{\text{LW}} (g; \delta g , \delta _{\xi} g),
\end{equation}
is conserved off-shell for the asymptotically field-independent Killing vector fields as well as field-independent Killing vector fields admitted by spacetime everywhere. For the case in which $\xi$ depends on the dynamical fields, it seems to be sensible replacing $\delta$ and the Lee-Wald symplectic current by $\hat{\delta}$ and $\omega ^{\mu} (g; \hat{\delta} g , \delta _{\xi} g)$ in Eq.\eqref{32}, respectively \cite{9}.\\
Similarly, in the Einstein-Maxwell theory, we can define the extended off-shell ADT current as
\begin{equation}\label{33}
   \mathfrak{J}^{\mu} _{\text{ADT}} (\Phi , \hat{\delta} \Phi ; \chi) =  \mathcal{J}^{\mu} _{\text{ADT}} (\Phi , \hat{\delta} \Phi ; \chi) + \frac{1}{2 \sqrt{-g}} \omega ^{\mu} (\Phi; \hat{\delta} \Phi , \delta _{\chi} \Phi).
\end{equation}
It is clear that the extended off-shell ADT current $\mathfrak{J}^{\mu} _{\text{ADT}}$ is conserved off-shell for the asymptotically symmetry generator $\chi$. By using Eq.\eqref{33}, the equation \eqref{25} can be written as
\begin{equation}\label{34}
   \sqrt{-g} \mathfrak{J}^{\mu} _{\text{ADT}} (\Phi , \hat{\delta} \Phi ; \chi) = \partial _{\nu} \left[ \sqrt{-g} \mathcal{Q} _{\text{ADT}} ^{\nu \mu} (\Phi , \hat{\delta} \Phi ; \chi) \right],
\end{equation}
where $\mathcal{Q} _{\text{ADT}} ^{\mu \nu} (\Phi , \hat{\delta} \Phi ; \chi)$ is defined as the extended off-shell ADT charge corresponding to the asymptotically symmetry generator $\chi$ and it is given by
\begin{equation}\label{35}
  \sqrt{-g} \mathcal{Q}_{\text{ADT}} ^{\mu \nu} (\Phi , \hat{\delta} \Phi ; \chi) = \frac{1}{2}\hat{\delta} K^{\mu \nu}(\Phi ; \chi) - \frac{1}{2} K^{\mu \nu}(\Phi ; \hat{\delta} \chi)- \xi ^{[\mu} \Theta ^{\nu]}(\Phi ; \hat{\delta} \Phi ).
\end{equation}
By defining $K^{\mu \nu}= \sqrt{-g} \tilde{K}^{\mu \nu}$ and $\Theta ^{\mu}= \sqrt{-g} \tilde{\Theta }^{\mu}$, the equation \eqref{35} becomes
\begin{equation}\label{36}
\begin{split}
   \mathcal{Q}_{\text{ADT}} ^{\mu \nu} (\Phi , \hat{\delta} \Phi ; \chi) = & \frac{1}{2}\hat{\delta} \tilde{K}^{\mu \nu}(\Phi ; \chi) + \frac{1}{4} g^{\alpha \beta} \hat{\delta} g_{\alpha \beta} \tilde{K}^{\mu \nu}(\Phi ; \chi) \\
     & - \frac{1}{2} \tilde{K}^{\mu \nu}(\Phi ; \hat{\delta} \chi)- \xi ^{[\mu} \tilde{\Theta} ^{\nu]}(\Phi ; \hat{\delta} \Phi ).
\end{split}
\end{equation}
By substituting Eq.\eqref{6} and Eq.\eqref{21} into Eq.\eqref{36}, we have
\begin{equation}\label{37}
  \mathcal{Q}_{\text{ADT}} ^{\mu \nu} (\Phi , \hat{\delta} \Phi ; \chi) = \mathcal{Q}_{(g)} ^{\mu \nu} (\Phi , \hat{\delta} \Phi ; \xi) +\mathcal{Q}_{(A)} ^{\mu \nu} (\Phi , \hat{\delta} \Phi ; \chi),
\end{equation}
where
\begin{equation}\label{38}
\begin{split}
   \mathcal{Q}_{(g)} ^{\mu \nu} (\Phi , \hat{\delta} \Phi ; \xi) = & - h^{\lambda [ \mu} \nabla _{\lambda} \xi ^{\nu]} + \xi ^{\lambda} \nabla ^{[\mu} h^{\nu]}_{\lambda} + \frac{1}{2} h \nabla ^{[\mu} \xi ^{\nu]} \\
     & - \xi ^{[\mu} \nabla _{\lambda} h^{\nu] \lambda} + \xi ^{[\mu} \nabla^{\nu]}h
\end{split}
\end{equation}
is the contribution from the gravity part and
\begin{equation}\label{39}
  \mathcal{Q}_{(A)} ^{\mu \nu} (\Phi , \hat{\delta} \Phi ; \chi)=2 \kappa \xi^{[\mu} F^{\nu] \alpha} \hat{\delta} A_{\alpha} +\kappa \lambda \left( \hat{\delta} F^{\mu \nu} + \frac{1}{2} h F^{\mu \nu} \right)
\end{equation}
is the contribution from the $U(1)$ gauge field part. In equations \eqref{38} and \eqref{39}, we have used the definition $h_{\mu \nu} = \hat{\delta} g_{\mu \nu}$. Now, we can define the perturbation of conserved charge by integrating from the extended off-shell ADT charge over a spacelike codimension two surface
\begin{equation}\label{40}
  \hat{\delta} Q(\chi) = c \int _{\Sigma} (d^{D-2} x) _{\mu \nu} \sqrt{-g} \mathcal{Q}_{\text{ADT}} ^{\mu \nu} (\Phi , \hat{\delta} \Phi ; \chi),
\end{equation}
where
\begin{equation}\label{41}
  (d^{D-2} x) _{\mu \nu} = \frac{1}{2(D-2)!} \varepsilon _{\mu \nu \alpha _{1} \cdots \alpha _{D-2}} dx^{\alpha_{1}} \cdots dx^{\alpha_{D-2}}
\end{equation}
and $c$ is just a universal constant. The charge defined by Eq.\eqref{37} is conserved off-shell for the asymptotically
 field-dependent symmetry generator $\chi$. If we set $\chi=(0,\lambda)$ then the conserved charge corresponding to the
 gauge generator $\lambda$ is just the electric charge. To find the conserved charge corresponds to a Killing
 vector field $\xi$ we should turn off gauge generator $\lambda$.
\section{Conserved charges of asymptotically AdS$_{3}$ spacetimes in the Einstein-Maxwell theory}
In this section, we consider the fall-off conditions presented in \cite{10} and we will try to obtain the conserved charges of spacetimes that obey the considered fall-off conditions. Assume that $\Lambda = - l^{-2}$, where $l$ is AdS radii. Let $r$ and $x^{\pm}= t/l \pm \phi$ are radial coordinate and the null coordinates, respectively.
\subsection{Asymptotic fall of conditions}
Now, we summarize the fall-off conditions presented in \cite{10}. The authors in \cite{10} have proposed the following fall-off conditions for asymptotically AdS$_{3}$ spacetimes in the Einstein-Maxwell theory (see \cite{3',4'} for another asymptotically  AdS$_{3}$ conditions in the context of the Einstein-Maxwell theory)
\begin{equation}\label{42}
  \begin{split}
     g_{\pm \pm} = & \frac{\kappa l^2}{4 \pi ^{2}} q_{\pm}^{2} \ln \left( \frac{r}{r_{0}}\right) + f_{\pm \pm} + \mathcal{O} \left( r^{-1} \ln r \right), \\
     g_{+-} = & -\frac{r^2}{2} + f _{+-}+ \mathcal{O} \left( r^{-1} \ln r \right),\\
     g_{rr} = & \frac{l^{2}}{r^{2}} + \frac{f_{rr}}{r^{4}} + \mathcal{O} \left( r^{-5} \ln r \right), \\
     g_{r \pm} = & \mathcal{O} \left( r^{-3} \ln r \right),
  \end{split}
\end{equation}
and
\begin{equation}\label{43}
  \begin{split}
     A_{\pm} =& -\frac{l}{2 \pi} q_{\pm} \ln \left( \frac{r}{r_{0}}\right) + \varphi _{\pm} + \mathcal{O} \left( r^{-2} \ln r \right), \\
      A_{r} =& \mathcal{O} \left( r^{-3} \ln r \right),
  \end{split}
\end{equation}
where $f_{\pm \pm} $, $f_{+-} $, $f_{rr} $, $q_{\pm} $ and $\varphi_{\pm} $ are arbitrary functions of the null coordinates $x^{\pm}$. The variation generated by the following symmetry generator $\chi$ preserves the fall-off conditions \eqref{42} and \eqref{43}
\begin{equation}\label{44}
  \begin{split}
     \xi^{\pm} = & T^{\pm} + \frac{l^{2}}{2r^{2}} \partial _{\mp}^{2} T^{\mp} + \mathcal{O} \left( r^{-4} \ln r \right),\\
     \xi^{r} =  & -\frac{r}{2} \left( \partial_{+} T^{+} + \partial_{-} T^{-} \right)+ \mathcal{O} \left( r^{-1} \right) ,\\
         \end{split}
\end{equation}
\begin{equation}\label{44'}
                  \lambda = \lambda _{0} + \mathcal{O} \left( r^{-2} \ln r \right)
                \end{equation}
where $T^{\pm}= T^{\pm} (x^{\pm}) $ and $\lambda _{0}= \lambda _{0}(x^{+},x^{-}) $ are arbitrary functions.
It is clear that, in this case, $\chi$ is independent of the dynamical fields.\\
 Under the action of a generic asymptotic symmetry generator $\chi$ spanned by Eq.\eqref{44} and Eq.\eqref{44'}, the dynamical
 fields transform as
\begin{equation}\label{58}
  \begin{split}
     \delta _{\chi} f_{rr} = & \partial _{+} ( T^{+} f_{rr}) + \partial _{-} ( T^{-} f_{rr}),\\
       \delta _{\chi} f_{+-} = & \partial _{+} ( T^{+} f_{+-}) + \partial _{-} ( T^{-} f_{+-}), \\
       \delta _{\chi} q_{\pm} =& \partial _{\pm} ( T^{\pm} q_{\pm}) + T^{\mp} \partial _{\mp} q_{\pm}, \\
       \delta _{\chi} \varphi _{\pm} =& \partial _{\pm} ( T^{\pm} \varphi_{\pm}) + T^{\mp} \partial _{\mp} \varphi_{\pm} + \frac{l}{4 \pi} q_{\pm}(\partial_{+} T^{+} + \partial _{-} T^{-}) + \partial _{\pm} \lambda_{0}.
  \end{split}
\end{equation}
We emphasize that the action of a generic asymptotic symmetry generator $\chi$ is defined by Eq.\eqref{11}
 and Eq.\eqref{12}.
\subsection{Conserved charges}
Now, we simplify the perturbation of the conserved charge Eq.\eqref{40} in the considered coordinates system. Since we consider the Einstein-Maxwell theory in 3-dimensions then Eq.\eqref{40} can be written as
\begin{equation}\label{45}
  \hat{\delta} Q(\chi) = -\frac{1}{2 \kappa} \int _{\Sigma} \sqrt{-g} \varepsilon_{\mu \nu \lambda} \mathcal{Q}_{\text{ADT}} ^{\mu \nu} (\Phi , \hat{\delta} \Phi ; \chi) dx^{\lambda},
\end{equation}
where we set $c= - \kappa ^{-1}$. We take the codimension two surface $\Sigma$ to be a circle with a radius of infinity so the Eq.\eqref{45} becomes
\begin{equation}\label{46}
  \hat{\delta} Q(\chi) = \frac{1}{ \kappa} \lim_{r \rightarrow \infty } \int _{0}^{2 \pi} \sqrt{-g} \left( \mathcal{Q}_{\text{ADT}} ^{r+} + \mathcal{Q}_{\text{ADT}} ^{r-} \right) d \phi .
\end{equation}
Hence, only two components of the off-shell ADT charge is important, i.e. we need to have $\mathcal{Q}_{\text{ADT}} ^{r+}$ and $ \mathcal{Q}_{\text{ADT}} ^{r-}$. By substituting Eq.\eqref{42}, Eq.\eqref{43}, \eqref{44} and \eqref{44'} into the equations \eqref{38} and \eqref{39}, we have
\begin{equation}\label{47}
\begin{split}
   \mathcal{Q}_{(g)} ^{r \pm} (\Phi , \hat{\delta} \Phi ; \xi) = & \frac{T^{\pm}}{2 l^{2} r} \hat{\delta} \left[ \frac{ f_{rr}}{l^{2}} - 4 f_{+-} \right]  \\
     & + \frac{T^{\mp}}{2 r} \hat{\delta} \left[ - \frac{\kappa}{\pi ^{2}} q_{\mp} ^{2} + \frac{\kappa}{\pi ^{2}} q_{\mp} ^{2} \ln\left( \frac{r}{r_{0}}\right) + \frac{4 f_{\mp \mp}}{l^{2}} \right]+ \mathcal{O}(r^{-2} \ln r),
\end{split}
\end{equation}
\begin{equation}\label{48}
\begin{split}
   \mathcal{Q}_{(A)} ^{\mu \nu} (\Phi , \hat{\delta} \Phi ; \chi)= & \frac{T^{\pm}}{2 r} \hat{\delta} \left[ \frac{\kappa}{\pi ^{2}} q_{+} q_{-} \ln\left( \frac{r}{r_{0}}\right) \right]- \frac{ \kappa T^{\pm}}{\pi l r} \left[ q_{-} \hat{\delta} \varphi _{+} + q_{+} \hat{\delta} \varphi _{-} \right] \\
     & + \frac{\kappa}{\pi l r} \lambda _{0} \hat{\delta} q_{\mp} + \mathcal{O}(r^{-2} \ln r).
\end{split}
\end{equation}
By substituting Eq.\eqref{47} and \eqref{48} into Eq.\eqref{46}, we find that
\begin{equation}\label{49}
  \begin{split}
     \hat{\delta} Q(\chi) = \hat{\delta} \int_{0}^{2 \pi} d \phi & \biggl\{ T^{+} \biggl[ \frac{1}{4 \kappa l} \left( \frac{ f_{rr}}{l^{2}} - 4 f_{+-} \right) + \frac{f_{++}}{\kappa l } - \frac{l}{8 \pi ^2 } q_{+}^2    \\
       &  + \frac{l}{4 \pi ^{2}} q_{+} \left( q_{+} + q_{-} \right) \ln \left( \frac{r}{r_{0}}  \right) - \frac{1}{2 \pi} \left( q_{-} \varphi _{+} + q_{+} \varphi _{-}\right) \biggr] \\
       & + T^{-} \biggl[ \frac{1}{4 \kappa l} \left( \frac{ f_{rr}}{l^{2}} - 4 f_{+-} \right) + \frac{f_{--}}{\kappa l } - \frac{l}{8 \pi ^2 } q_{-}^2    \\
       & + \frac{l}{4 \pi ^{2}} q_{-} \left( q_{+} + q_{-} \right) \ln \left( \frac{r}{r_{0}}  \right) - \frac{1}{2 \pi} \left( q_{-} \varphi _{+} + q_{+} \varphi _{-}\right) \biggr] \\
       & + \frac{1}{2 \pi} \lambda_{0} \left( q_{+} + q_{-} \right) \biggr\} \\
       + \frac{1}{2\pi} \int_{0}^{2 \pi} & d \phi \left( T^{+}+T^{-} \right) \left( \varphi _{+} \hat{\delta} q_{-} + \varphi _{-} \hat{\delta} q_{+} \right).
  \end{split}
\end{equation}
The last term in Eq.\eqref{49} is the non-integrable part of the conserved charge perturbation which corresponds to
the symmetry generator $\chi$.
 As we mentioned earlier, by setting $\xi=0$, or equivalently $T^{\pm}=0$, one finds the conserved charge
 corresponding to the pure $U(1)$ gauge symmetry
\begin{equation}\label{52}
  Q(\lambda)=\frac{1}{2 \pi} \int_{0}^{2 \pi} \lambda_{0} \left( q_{+} + q_{-} \right) d \phi.
\end{equation}
Also, by setting $\lambda=0$, we find the following expression for the conserved charge corresponding
to the diffeomorphism generator $\xi$
\begin{equation}\label{70}
  \begin{split}
     \hat{\delta} Q(\xi) = \hat{\delta} \int_{0}^{2 \pi} d \phi & \biggl\{ T^{+} \biggl[ \frac{1}{4 \kappa l} \left( \frac{ f_{rr}}{l^{2}} - 4 f_{+-} \right) + \frac{f_{++}}{\kappa l } - \frac{l}{8 \pi ^2 } q_{+}^2    \\
       &  + \frac{l}{4 \pi ^{2}} q_{+} \left( q_{+} + q_{-} \right) \ln \left( \frac{r}{r_{0}}  \right) - \frac{1}{2 \pi} \left( q_{-} \varphi _{+} + q_{+} \varphi _{-}\right) \biggr] \\
       & + T^{-} \biggl[ \frac{1}{4 \kappa l} \left( \frac{ f_{rr}}{l^{2}} - 4 f_{+-} \right) + \frac{f_{--}}{\kappa l } - \frac{l}{8 \pi ^2 } q_{-}^2    \\
       & + \frac{l}{4 \pi ^{2}} q_{-} \left( q_{+} + q_{-} \right) \ln \left( \frac{r}{r_{0}}  \right) - \frac{1}{2 \pi} \left( q_{-} \varphi _{+} + q_{+} \varphi _{-}\right) \biggr] \biggr\}\\
       + \frac{1}{2\pi} \int_{0}^{2 \pi} & d \phi \left( T^{+}+T^{-} \right) \left( \varphi _{+} \hat{\delta} q_{-} + \varphi _{-} \hat{\delta} q_{+} \right).
  \end{split}
\end{equation}
Due to the presence of the logarithmic terms in Eq.\eqref{70}, if one considers just diffeomorphism, i.e. one
 sets $\lambda$ to be zero, the expression for conserved charge perturbation
 (corresponding to the asymptotically Killing vector $\xi$) diverges at spatial infinity. Hence, the ordinary
 quasi-local conserved charge method presented in \cite{1,3} fails to give the finite charges in the 3D Einstein-Maxwell
 theory. To avoid this problem, we must consider both $\xi$ and $\lambda =  \lambda _{\xi}$ together. Therefore, we consider the expression in Eq.\eqref{40} for the conserved charge such that it just corresponds to diffeomorphism generator $\xi$. In this way, we have a transformation such that it is just generated by a vector field, i.e. $\chi=(\xi, \lambda _{\xi}) \rightarrow \xi$. The subscript $\xi$ in $\lambda _{\xi}$ indicates that $\lambda _{\xi}$ is a function of $\xi$ and it is not an independent symmetry generator. We remind that the boundary conditions Eq.\eqref{42} and Eq.\eqref{43} are preserved under diffeomorphism generated by the asymptotically Killing vector field Eq.\eqref{44}. By substituting equations \eqref{42},\eqref{43}, \eqref{44} and $\lambda = \lambda _{\xi}$ into Eq.\eqref{40} along with $\chi=(\xi, \lambda _{\xi}) \rightarrow \xi$ and $Q(\chi) \rightarrow Q^{\prime}(\xi) $, we have
 \begin{equation}\label{72}
  \begin{split}
     \hat{\delta} Q^{\prime}(\xi) = \hat{\delta} \int_{0}^{2 \pi} d \phi & \biggl\{ T^{+} \biggl[ \frac{1}{4 \kappa l} \left( \frac{ f_{rr}}{l^{2}} - 4 f_{+-} \right) + \frac{f_{++}}{\kappa l } - \frac{l}{8 \pi ^2 } q_{+}^2    \\
       &  + \frac{l}{4 \pi ^{2}} q_{+} \left( q_{+} + q_{-} \right) \ln \left( \frac{r}{r_{0}}  \right) - \frac{1}{2 \pi} \left( q_{-} \varphi _{+} + q_{+} \varphi _{-}\right) \biggr] \\
       & + T^{-} \biggl[ \frac{1}{4 \kappa l} \left( \frac{ f_{rr}}{l^{2}} - 4 f_{+-} \right) + \frac{f_{--}}{\kappa l } - \frac{l}{8 \pi ^2 } q_{-}^2    \\
       & + \frac{l}{4 \pi ^{2}} q_{-} \left( q_{+} + q_{-} \right) \ln \left( \frac{r}{r_{0}}  \right) - \frac{1}{2 \pi} \left( q_{-} \varphi _{+} + q_{+} \varphi _{-}\right) \biggr] \\
       & + \frac{1}{2 \pi} \lambda_{\xi} \left( q_{+} + q_{-} \right) \biggr\} \\
       + \frac{1}{2\pi} \int_{0}^{2 \pi} & d \phi \left( T^{+}+T^{-} \right) \left( \varphi _{+} \hat{\delta} q_{-} + \varphi _{-} \hat{\delta} q_{+} \right).
  \end{split}
\end{equation}
Now, we fix the $U(1)$ gauge $\lambda_{\xi}$ such that the logarithmic terms appeared in Eq.\eqref{72} to be removed. So we set $\lambda_{\xi}$ as follows:\footnote{This choice for $\lambda _{\xi}$ is different from Eq.\eqref{44'}, which is presented in subsection 3.1. It does not make a problem because it does not destroy the boundary conditions Eq.\eqref{43}. One can use Eq.\eqref{12} to check that, under pure gauge transformation generated by the gauge parameter Eq.\eqref{50}, the boundary conditions Eq.\eqref{43} transforms like $ A_{\pm} \rightarrow -\frac{l}{2 \pi} q^{\prime}_{\pm} \ln \left( \frac{r}{r_{0}}\right) + \varphi^{\prime} _{\pm} + \mathcal{O} \left( r^{-2} \ln r \right)$.}
\begin{equation}\label{50}
\begin{split}
   \lambda _{\xi}  & = \xi^{\mu} A_{\mu} \\
     & = -\frac{l}{2 \pi} \left( q_{+}T^{+} + q_{-}T^{-} \right) \ln \left( \frac{r}{r_{0}}  \right) + \left( \varphi_{+}T^{+} + \varphi_{-}T^{-} \right) + \mathcal{O}\left( r^{-2} \ln r \right).
\end{split}
\end{equation}
By substituting Eq.\eqref{50} into Eq.\eqref{72}, one finds the following expression for the conserved charge perturbation corresponding to the asymptotically Killing vector $\xi$
\begin{equation}\label{51}
  \begin{split}
     \hat{\delta} Q^{\prime}(\xi) = \hat{\delta} \int_{0}^{2 \pi} d \phi & \biggl\{ T^{+} \biggl[ \frac{1}{4 \kappa l} \left( \frac{ f_{rr}}{l^{2}} - 4 f_{+-} \right) + \frac{f_{++}}{\kappa l } - \frac{l}{8 \pi ^2 } q_{+}^2    \\
       & + \frac{1}{2 \pi} q_{+} \left( \varphi _{+} - \varphi _{-}\right) \biggr] \\
       & + T^{-} \biggl[ \frac{1}{4 \kappa l} \left( \frac{ f_{rr}}{l^{2}} - 4 f_{+-} \right) + \frac{f_{--}}{\kappa l } - \frac{l}{8 \pi ^2 } q_{-}^2    \\
       & - \frac{1}{2 \pi} q_{-} \left( \varphi _{+} - \varphi _{-}\right) \biggr] \biggr\} \\
       + \frac{1}{2\pi} \int_{0}^{2 \pi} & d \phi \left( T^{+}+T^{-} \right) \left( \varphi _{+} \hat{\delta} q_{-} + \varphi _{-} \hat{\delta} q_{+} \right).
  \end{split}
\end{equation}
It is easy to see that, using this combined transformation, the logarithmic terms are removed.\\
 As one can see, the conserved charges, Eq.\eqref{70}, corresponding to the \textit{pure diffeomorphism} generated by the vector field in Eq.\eqref{44} diverges at spatial infinity. We emphasize that the $U(1)$ gauge transformations are suppressed when we consider the \textit{pure diffeomorphism} . To remove the divergent term, we have introduced the combined transformation $\chi=(\xi, \lambda_{\xi})$, where $\lambda_{\xi}=\xi_{\mu}A^{\mu}$ is a $U(1)$ gauge transformation generator. Conserved charge corresponding to the \textit{pure $U(1)$ transformations} generated by $\lambda_{\xi}=\xi_{\mu}A^{\mu}$, where $\xi$ is given by Eq.\eqref{44}, also diverges at spatial infinity. By combining the \textit{pure diffeomorphism} generated by $\xi$  and a \textit{pure $U(1)$ transformation} generated by $\lambda_{\xi}=\xi_{\mu}A^{\mu}$, the divergent terms appeared in the combined conserved charge cancel each other out, then we find a finite conserved charge. Since the combined transformation generated by $\chi=(\xi, \xi_{\mu}A^{\mu})$ depends on just the pure diffeomorphism generator $\xi$, then the conserved charge corresponding to $\chi$ is just conserved charge corresponding to the diffeomorphism generator $\xi$. As we have mentioned earlier, the boundary conditions, Eq.\eqref{42} and Eq.\eqref{43}, are preserved under the diffeomorphism generated by the vector field, Eq.\eqref{44}, and the $U(1)$ gauge transformation generated by $\lambda_{\xi}=\xi_{\mu}A^{\mu}$. Thus we were able to obtain finite conserved charge, Eq.\eqref{51}, associated with $\xi$. In parallel with the paper \cite{10}, a general transformation can be generated by two generators $ (\xi,\eta+\xi_{\mu}A^{\mu})$, where $\xi$ and $\eta$ are generators of the diffeomorphism and the $U(1)$ gauge transformations. By suppressing the $U(1)$ gauge transformations, that is $\eta=0$, one finds the conserved charge corresponding to the diffeomorphism. This case is indeed what we have used to find the conserved charge corresponding to the transformation generated by $\chi=(\xi,\xi_{\mu}A^{\mu})$. In contrast, by suppressing the diffeomorphism, that is $\xi=0$, one finds the conserved charges corresponding to the $U(1)$ gauge transformations. In this case, the appropriate $U(1)$ gauge transformation generator which preserves boundary conditions \eqref{42} and \eqref{43} is given by \eqref{44'} (set $\eta=\lambda$) and the corresponding conserved charge is \eqref{52}. In the next subsection, we will consider the field equations and integrability condition.
\subsection{On-shell case and integrability condition}
By substituting Eq.\eqref{42} and Eq.\eqref{43} into the field equations \eqref{4} and \eqref{5}, we have
\begin{equation}\label{53}
  \begin{split}
       & \mathcal{E}_{(g)}^{rr}= \frac{1}{ l^{4}} \left( \frac{ f_{rr}}{l^{2}} - 4 f_{+-} - \frac{\kappa l^{2}}{2 \pi^{2}} q_{+} q_{-}\right) + \mathcal{O}(r^{-2}) \\
       & \mathcal{E}_{(g)}^{r \pm}= \mathcal{O}(r^{-3}) ,\hspace{0.7 cm} \mathcal{E}_{(g)}^{\pm \pm}= \mathcal{O}(r^{-6}),\hspace{0.7 cm} \mathcal{E}_{(g)}^{+-}= \mathcal{O}(r^{-6}),\\
       & \mathcal{E}_{(A)}^{r}= \mathcal{O}(r^{-1}) ,\hspace{0.7 cm} \mathcal{E}_{(A)}^{\pm}= \mathcal{O}(r^{-4}).
  \end{split}
\end{equation}
It is clear that, at spatial infinity, these field equations satisfy when
\begin{equation}\label{54}
  \frac{ f_{rr}}{l^{2}} - 4 f_{+-} = \frac{\kappa l^{2}}{2 \pi^{2}} q_{+} q_{-}.
\end{equation}
If one assumes that $\varphi _{\pm}$ are functions of $q_{+}$ and $q_{-}$ then the integrability condition, $\hat{\delta}_{[1} \hat{\delta}_{2]} Q(\xi) =0$, leads to \cite{10}
\begin{equation}\label{55}
  \varphi _{\pm} = \frac{1}{2}\frac{\hat{\delta}\mathcal{V}}{\hat{\delta} q_{\mp}},
\end{equation}
where $\mathcal{V}=\mathcal{V}(q_{+},q_{-})$. By substituting Eq.\eqref{54} and Eq.\eqref{55} into Eq.\eqref{51}, we find the following expression for the conserved charge corresponding to the asymptotically Killing vector $\xi$
\begin{equation}\label{56}
Q^{\prime}(\xi) = Q^{\prime}_{\xi}(T^{+}) + Q^{\prime}_{\xi}(T^{-}),
\end{equation}
where
\begin{equation}\label{57}
Q^{\prime}(T^{\pm})= \int_{0}^{2 \pi} d \phi T^{\pm} \left[ \frac{f_{\pm \pm}}{\kappa l } \mp \frac{l}{8 \pi ^2 } q_{\pm} \left( q_{+} - q_{-} \right) \pm \frac{1}{2 \pi} q_{\pm} \left( \varphi _{+} - \varphi _{-}\right) + \frac{\mathcal{V}}{4\pi} \right].
\end{equation}
The results obtained in this section, the conserved charge Eq.\eqref{52} corresponds to the pure $U(1)$ gauge symmetry and the conserved charge Eq.\eqref{56} corresponds to the asymptotically Killing vector $\xi$, are consistent with the results of \cite{10}.

\section{Conclusion}
In this paper we have considered the Einstein-Maxwell theory which is described by the Lagrangian introduced in Eq.\eqref{1}.
 We have defined a combined transformation made up of diffeomorphism and $U(1)$ gauge transformation. We have denoted
the generator of such transformations by $\chi= (\xi,\lambda)$, where $\xi$ is the diffeomorphism generator vector field
 and $\lambda$ is the generator of $U(1)$ gauge transformations. The metric and the $U(1)$ gauge field under
 transformation generated by $\chi$ are transformed as Eq.\eqref{11} and Eq.\eqref{12}. To have a general discussion, we have assumed
 that $\chi$ is field dependent. We have defined the extended off-shell ADT current in Eq.\eqref{33} which is
 conserved off-shell for the asymptotically field-dependent symmetry generator $\chi$. We have used the extended
 off-shell current, Eq.\eqref{33}, to define the extended off-shell ADT charge, Eq.\eqref{37}. Then, by integrating
 the extended off-shell ADT charge over a spacelike codimension-2 surface, we have defined the conserved charge
 perturbation, Eq.\eqref{40}, which corresponds to the asymptotically field-dependent symmetry generator $\chi$. In section 3, we
 have considered the Einstein-Maxwell theory in 3 dimensions. The fall-off conditions for the asymptotically AdS$_{3}$
 spacetimes are given by Eq.\eqref{42} and Eq.\eqref{43}. The considered fall-off conditions are preserved by
 transformations whose generators are given by Eq.\eqref{44} and Eq.\eqref{44'}. We have found the conserved charge perturbation, Eq.\eqref{49}, of spacetimes, which obeys the fall-off condition Eq.\eqref{42}, corresponds to the symmetry generator $\chi$.
 It is clear that the obtained conserved charge perturbation, Eq.\eqref{49}, is not integrable. The conserved charge
 perturbation which corresponds to the asymptotically Killing vector $\xi$ ($\lambda_{0}=0$) diverges at spatial infinity
 (see Eq.\eqref{70}). To avoid this problem, we have considered both $\xi$ and
 $\lambda =  \lambda _{\xi}$ together. We have considered the
 expression in Eq.\eqref{40} for the conserved charge such that it is just corresponds to the diffeomorphism generator $\xi$.
 In this way, we have a transformation such that it just generated by a vector field,
 i.e. $\chi=(\xi, \lambda _{\xi}) \rightarrow \xi$. We have defined the conserved charge $Q^{\prime}(\xi)$
 corresponding to the combined transformation $(\xi, \lambda _{\xi})$. By gauge fixing
 $\lambda _{\xi} = \xi^{\nu} A_{\nu}$, we have obtained a finite conserved charge perturbation
 corresponding to the asymptotically Killing vector field $\xi$, Eq.\eqref{44}, at spatial infinity (see Eq.\eqref{51}).
 Also, we have obtained the conserved charge in Eq.\eqref{52} corresponding to the $U(1)$ gauge symmetry. One can
 solve the field equations Eq.\eqref{53} asymptotically when Eq.\eqref{54} holds.
 We have assumed that $\varphi_{\pm}=\varphi_{\pm} (q_{+},q_{-})$ and used the integrability condition to
 simplify the expression Eq.\eqref{51}. Then we have found the expression Eq.\eqref{56} for the conserved charge corresponding
 to the asymptotically Killing vector $\xi$. The results obtained by using quasi-local method presented in this paper
 (the conserved charge Eq.\eqref{52} corresponds to the pure $U(1)$ gauge symmetry and the conserved charge Eq.\eqref{56}
 corresponds to the asymptotically Killing vector $\xi$) are consistent with those of \cite{10}, where the
 authors have used the Hamiltonian formalism to find the corresponding results.
 \section{Acknowledgments}
M. R. Setare  thanks Dr. A. Sorouri for his help in improvement the English of the text.

\end{document}